\documentclass[conference]{IEEEtran}
\IEEEoverridecommandlockouts

\usepackage[utf8]{inputenc}
\usepackage{booktabs}
\usepackage{amsmath}
\usepackage{xcolor}
\usepackage{tikz}
\usepackage{url}
\usepackage[numbers,sort&compress]{natbib}
\usetikzlibrary{arrows.meta,positioning,fit}

\title{NepTTS-Bench: Native-Speaker Benchmarking and NepaliMOS for Nepali Text-to-Speech}

\author{
\IEEEauthorblockN{Ashish Thapa}
\IEEEauthorblockA{Ampixa Labs\
Email: ashish@ampixa.com}
\and
\IEEEauthorblockN{Rupak Ghimire}
\IEEEauthorblockA{Kathmandu University\
Email: rupakghimire@gmail.com}
\and
\IEEEauthorblockN{Lekhnath Sharma Pathak}
\IEEEauthorblockA{Central Department of Linguistics\
Tribhuvan University\
Email: lekhnath.pathak@cdl.tu.edu.np}
}

\begin{document}
\maketitle

\begin{abstract}
Nepali TTS has reached a point where multiple systems are available, but there is no shared evaluation to compare them. NepTTS-Bench was created to address what existing automated and manual evaluation systems lack for Nepali TTS. A controlled benchmark with phonological grounding which contains 205 sentences (77 minimal pairs targeting aspiration, retroflex/dental contrasts, nasalization, gemination, and schwa deletion) which is rated by 180 speakers. There are 6,962 ratings for 10 TTS configurations and two human references. Along with this 7 ASR systems were evaluated for round-trip CER. SCOREQ was also run as an automated baseline. The key finding is that automated evaluation like SCOREQ prefers proprietary systems like ElevenLabs but native listeners prefer Piper. This is contributed by the fact that most of the systems not properly tuned for Nepali tend to show Hindi phonological behaviour and our minimal pairs are designed to surface such contrasts. We also present NepaliMOS predictor which is fine tuned on collected ratings. NepaliMOS achieves system-level $\rho = 0.90$ on the dense panel but drops to utterance-level $\rho = 0.43$ under leave-one-system-out evaluation. This is statistically significant and confirms that NepaliMOS captures genuine signal but cross-provider transfer is limited. A human calibration is recommended for new systems for which we release a Python evaluation package for practitioners to run the benchmark against new systems.
\end{abstract}

\section{Introduction}

Nepali TTS are available through open models and commercial APIs, but we don't have a proper framework to compare them on nepali prompts based on native judgements. 
Systems designed for English do not transfer cleanly, and published evaluations have never shared a common test set, a common listener pool, or a common protocol. 
Among the different papers examined, the numbers don't line up. There is no shared benchmark that captures language-specific pronunciation errors with native-speaker ratings attached. 
MOS depends on its protocol \citep{cooper2024synthetic_eval}. The listener pool, interface design, and test material all shift the result. Without a common setting, it is not possible to tell whether a gap between two reported scores is real or an artifact of how the studies were run.

NepTTS-Bench is our answer to this. It has 205 Nepali sentences, 77 of which are minimal pairs targeting aspiration, retroflex and dental contrasts, nasalization, gemination, and schwa deletion. The remaining 58 cover prosodic variation, homographs, code-mixed text, and other cases that matter in practice. Audio from 40 named systems or voices is rated under one controlled protocol by 180 native Nepali speakers. An evaluation CLI and rating web app will be released with the data upon acceptance. Which signals can actually be trusted for Nepali TTS evaluation, and where do they break down?

NepTTS-Bench is the infrastructure for the whole judgement pipeline. The contribution here is the evaluation setting: phonologically motivated prompts, native-speaker MOS under one protocol across proprietary, open, and reference conditions, an analysis of where automated metrics diverge from native-speaker judgment, and NepaliMOS — a Nepali-supervised MOS predictor with its practical limits tested and reported directly.

\section{Related Work}

MOS is the standard subjective measure for TTS naturalness and quality \citep{streijl2016mos, cooper2024synthetic_eval}. 
Shared tasks such as VoiceMOS encouraged automated MOS prediction with systems such as UTMOS \citep{huang2022voicemos, saeki2022utmos}. 
Recent no-reference quality models, including SCOREQ \citep{ragano2024scoreq} and DNSMOS \citep{reddy2022dnsmos}, are useful general tools, but their validation is dominated by higher-resource settings. 
ASR round-trip evaluation is a complementary intelligibility proxy; TTSDS2 shows that ASR and metric choice can change TTS rankings \citep{minixhofer2025ttsds}. 
For Nepali, this prior work suggests that metric behavior should be reported as evidence. 

Nepali TTS has moved from rule-based and concatenative work to neural systems. Early systems include ESNOLA synthesis \citep{chettri2013esnola}, TD-PSOLA work \citep{malla2015tdpsola}, and FreeTTS/formant-style approaches \citep{shah2018freetts}. 
The Bhasa Sanchar/Festival line focused on unit selection, corpus design, and preprocessing/postprocessing for quality \citep{ghimire2017enhancing, bajracharya2018natural}. Open low-resource voice-building efforts also included Nepali \citep{sodimana2018step}. 
Recent Nepali work uses Tacotron2, FastPitch/HiFi-GAN, FastSpeech/MelGAN, vocoder comparisons, and few-shot voice cloning \citep{dhakal2023tacotron2, dongol2023transformer, lamichhane2023fastspeech, dhakalchhetri2025vocoder, shrestha2026voicecloning}. 
These show that Nepali neural TTS is feasible, but they do not provide a reusable native-speaker benchmark across current proprietary and open systems.

Nepali speech resources and ASR systems provide useful infrastructure for round-trip evaluation. FLEURS covers Nepali in a multilingual speech benchmark \citep{conneau2022fleurs}, XLS-R checkpoints have been fine-tuned for Nepali ASR \citep{gagan2023xlsr}, and NepConformer gives a recent Conformer-based Nepali ASR baseline \citep{poudel2026nepconformer}. 
Our target is TTS benchmarking: controlled Nepali prompts with native MOS ratings, cross-category system comparison, and a language-matched quality predictor tested against the native ratings we collected.

\section{Methodology}

\subsection{Benchmark Dataset Preparation}

The benchmark contains 205 sentences informed by Nepali phonology \citep{khatiwada2009, pokharel1989, bandhu1971}. Its core is 147 sentences arranged into 77 minimal pairs. These pairs test contrasts that matter for intelligibility and perceived accent in Nepali TTS: laryngeal contrast across stop and affricate series, retroflex/dental opposition, oral/nasal vowel contrasts, gemination, and schwa deletion. The remaining 58 sentences cover a range of non-minimal-pair cases: question intonation, emotion, contrastive stress, homographs, code-mixed robustness sentences, tongue twisters, newspaper text, and common phrases.

Sentence preparation was an engineering audit rather than a formal phonological validation. Each minimal pair candidate was checked for grammaticality and spelling; we also verified that the target contrast was isolated and not confounded by other differences. Twelve pairs were replaced and six corrected. We treat the resulting set as phonologically motivated test material. It is not a validated account of the Nepali phonological system, and we do not claim it as one \citep{cooper2024synthetic_eval}.
\subsection{NepaliMOS Architecture}

NepaliMOS exists because we discovered early that running SCOREQ alone penalizes nothing on signal cleanliness without flagging pronunciation errors that native listeners immediately notice. SCOREQ training cannot detect vowels and aspiration in Hindi accents, but Nepali listeners capture them. The purpose of NepaliMOS is to test whether Nepali-specific supervision can recover those native speaker preferences better than general-purpose metrics. It is not a replacement for human evaluation on systems outside the training distribution.

\begin{figure*}[t]
\centering
\begin{tikzpicture}[
  font=\footnotesize,
  block/.style={
    draw,
    rounded corners=2pt,
    align=center,
    minimum height=0.76cm,
    text width=2.05cm,
    inner sep=3pt
  },
  note/.style={
    draw,
    rounded corners=2pt,
    align=center,
    minimum height=0.54cm,
    text width=2.45cm,
    inner sep=2pt,
    font=\scriptsize
  },
  lane/.style={font=\scriptsize\scshape, anchor=east},
  arrow/.style={-{Latex[length=2mm]}, thick},
  weakarrow/.style={-{Latex[length=2mm]}, thick, dashed}
]
\node[lane] at (-0.15,1.35) {Supervision};
\node[lane] at (-0.15,0) {Audio pathway};
\node[lane] at (-0.15,-2.25) {Evaluation scope};

\node[block, fill=gray!8] at (1.05,0) (wave) {Audio clip\\$x_i$};
\node[block, fill=cyan!7] at (3.55,0) (encoder) {IndicWav2Vec\\base encoder};
\node[block, fill=blue!7] at (6.05,0) (pool) {Masked mean\\pooling};
\node[block, fill=orange!11, text width=2.45cm] at (8.85,0) (head) {Regression head\\$768\!\rightarrow\!256\!\rightarrow\!1$\\ReLU, dropout 0.1};
\node[block, fill=green!9] at (11.85,0) (pred) {Predicted MOS\\$\hat{y}_i$};

\draw[arrow] (wave) -- (encoder);
\draw[arrow] (encoder) -- (pool);
\draw[arrow] (pool) -- (head);
\draw[arrow] (head) -- (pred);

\node[note, fill=cyan!3] at (3.55,-0.86) (freeze) {lower layers frozen\\top 4 layers trainable};
\node[note, fill=blue!3] at (6.05,-0.86) (embedding) {768-d clip\\representation};

\node[
  draw,
  dashed,
  rounded corners=3pt,
  inner xsep=7pt,
  inner ysep=8pt,
  fit=(wave)(encoder)(pool)(head)(pred)(freeze)(embedding),
  label={[font=\scriptsize, yshift=-1pt]below:Audio-only checkpoint: no reference text/ASR input}
] (modelbox) {};

\node[note, fill=yellow!13, text width=2.7cm] at (5.45,1.35) (ratings) {Native MOS ratings\\$r_{i1},\ldots,r_{im_i}$};
\node[note, fill=yellow!13, text width=2.45cm] at (8.45,1.35) (target) {Clip mean target\\$y_i=\frac{1}{m_i}\sum_j r_{ij}$};
\node[note, fill=yellow!13, text width=2.3cm] at (11.85,1.35) (loss) {MSE\\$(\hat{y}_i-y_i)^2$};
\draw[arrow] (ratings) -- (target);
\draw[arrow] (target) -- (loss);
\draw[weakarrow] (pred) -- (loss);

\node[note, fill=purple!7, text width=2.85cm] at (3.0,-2.55) (randomsplit) {90/10 utterance split\\within dense panel};
\node[note, fill=purple!7, text width=2.85cm] at (6.45,-2.55) (loso) {LOSO transfer check\\held-out TTS provider};
\node[note, fill=red!6, text width=3.25cm] at (10.35,-2.55) (scope) {Auxiliary score only\\human calibration for new providers};
\draw[arrow] (randomsplit) -- (loso);
\draw[arrow] (loso) -- (scope);
\end{tikzpicture}
\caption{NepaliMOS architecture and evaluation protocol. The implemented checkpoint is an audio-only MOS predictor: Nepali TTS audio is encoded with IndicWav2Vec, pooled to a clip representation, and mapped to a scalar MOS by a lightweight regression head. Supervision is the clip-level mean of native-speaker MOS ratings, optimized with MSE. The utterance split measures within-panel prediction, while leave-one-system-out (LOSO) validation tests transfer to systems unseen during training and motivates human calibration for new providers.}
\label{fig:nepalimos_architecture}
\end{figure*}
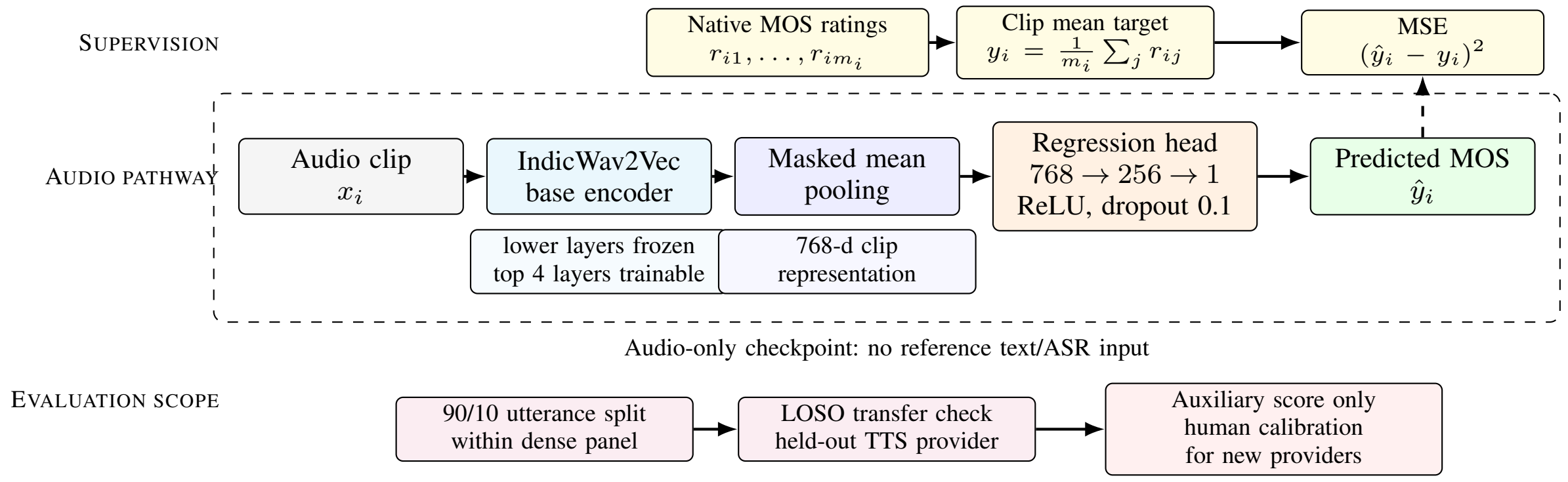

Figure~\ref{fig:nepalimos_architecture} summarizes the architecture and validation protocol.
The model fine-tunes IndicWav2Vec base \citep{javed2022indicwav2vec}. Audio is encoded by the SSL backbone, frame representations are mean-pooled, and a two-layer regression head maps $768 \rightarrow 256 \rightarrow 1$ with ReLU and dropout 0.1. The top four transformer layers are unfrozen and lower backbone layers remain frozen. The training target is the audio-level mean MOS label for each unique TTS clip, and training uses MSE loss. The final matched set contains 1,697 unique TTS clips aggregated from 5,865 matched-audio ratings, 5,804 of them from native-speaker registrations. The main checkpoint uses a stratified 90/10 utterance-level split, giving 1,530 training clips and 167 held-out validation clips. The prediction head has 197k parameters and the unfrozen backbone has 28.4M trainable parameters. We use differential learning rates ($10^{-4}$ for the head, $10^{-5}$ for the unfrozen backbone), dropout, batch size 2, and best-validation-loss checkpoint selection. To test whether NepaliMOS works on a provider it has not seen before, we also run leave-one-system-out validation: all clips from one TTS system are removed from training and used only for testing.

\section{Experiments}

The full deployed corpus contains audio for 40 named systems or voices. The main statistical panel is smaller by design: it uses the 10 TTS configurations with dense, comparable rating coverage, plus two human reference conditions. On the proprietary side, this panel covers ElevenLabs v3, Google Gemini Flash, Microsoft Edge TTS (two voices: Sagar and Hemkala), TingTing.io (three voices: Asmita, Subina, Sambriddhi), and gTTS. Piper is the open/local representative, evaluated as a VITS-based system. Human read speech and natural speech were blinded and included as references. Gemini required special handling: raw Devanagari input often produces Hindi-accented output, so we evaluated it in two modes and label the default-input variant ``Gemini (Hindi)'' to reflect this.

The remaining generated systems are pilot and screening material, not full statistical conditions. These span a wide range: SpeechT5 checkpoints from several contributors, eSpeak, MMS-TTS, VITS and VITS2 Nepali checkpoints, VoxCPM-family variants, multiple Parler-TTS Nepali fine-tunes, and internal voice experiments. Some have full prompt coverage but only sparse ratings; the newest have between two and 41 rated prompts. They are kept separate from the main MOS confidence intervals but remain part of the public release.

Native Nepali speakers rated blinded audio on a 1--5 MOS scale through a web application. Raters received balanced randomized samples across systems and never saw system identity. Same-sentence stimuli were spaced at least three positions apart. We report this setup because subjective synthetic-speech scores are protocol-dependent and should not be compared as if they were absolute system properties \citep{cooper2024synthetic_eval}.

ASR round-trip CER was computed with Google Chirp~2, Meta \texttt{mms-1b-all}, XLS-R Nepali, NepConformer CTC-BPE \citep{poudel2026nepconformer}, Whisper-small, Whisper large-v2, and Whisper large-v3-turbo. Whisper large-v3-turbo is retained only as a negative-finding run because of hallucination. SCOREQ is run as an external no-reference quality metric \citep{ragano2024scoreq}. Human MOS confidence intervals use 95\% cluster bootstrap over raters with 2,000 resamples. We report Spearman's $\rho$ for ranking agreement, Benjamini--Hochberg correction for automated-metric tests, and Krippendorff's $\alpha$ on the rater-by-sample matrix.

\section{Results and Discussion}
\label{sec:results}

\begin{table*}[t]
\centering
\caption{Human MOS ratings from 180 contributing native Nepali speakers. Confidence intervals are 95\% cluster-bootstrap intervals over raters.}
\label{tab:human_mos_slt}
\footnotesize
\setlength{\tabcolsep}{4pt}
\begin{tabular}{llccc@{\hspace{1.0em}}llccc}
\toprule
System & Access & MOS & 95\% CI & $n$ & System & Access & MOS & 95\% CI & $n$ \\
\midrule
Natural speech & Ref. & \textbf{3.95} & [3.79,4.15] & 597 &
Gemini (Hindi) & Prop. & 3.34 & [3.15,3.53] & 514 \\
Human (read) & Ref. & 3.64 & [3.44,3.82] & 552 &
Gemini Flash & Prop. & 3.25 & [2.90,3.68] & 618 \\
TingTing Asmita & Prop. & 3.55 & [3.27,3.88] & 651 &
TingTing Sambriddhi & Prop. & 3.20 & [2.91,3.55] & 617 \\
Piper & Open/local & 3.50 & [3.31,3.69] & 515 &
gTTS & Prop. & 2.60 & [2.38,2.83] & 510 \\
TingTing Subina & Prop. & 3.47 & [3.24,3.79] & 644 &
 & & & & \\
ElevenLabs v3 & Prop. & 3.47 & [3.28,3.66] & 516 &
 & & & & \\
Edge TTS Sagar & Prop. & 3.38 & [3.02,3.77] & 609 &
 & & & & \\
Edge TTS Hemkala & Prop. & 3.37 & [3.10,3.67] & 619 &
 & & & & \\
\bottomrule
\end{tabular}
\end{table*}

Table~\ref{tab:human_mos_slt} tells a cleaner story about the bottom than the top. gTTS sits clearly below every other system; that gap is real. What is less clear is any ordering within the top cluster, where the confidence intervals overlap for most systems and the data do not support naming a single winner. Nepali-specific, cloud, and open-source systems can all be competitive here. The low inter-rater agreement (Krippendorff's $\alpha=0.17$) is part of that result, not a flaw to set aside. It limits how precisely any protocol can rank systems that are close in quality \citep{cooper2024synthetic_eval}, which is relevant context for anyone interpreting these scores.

The metric disagreements are informative. A generic no-reference quality model and ASR-based intelligibility both reward clean, easy-to-transcribe audio. Native listeners care about more than that: accent, rhythm, and Nepali-specific phonological realizations all factor in, and those are not what SCOREQ or Whisper was trained to detect. Table~\ref{tab:auto_slt} shows NepaliMOS as the only metric that survives Benjamini-Hochberg correction at $n=9$ systems. Among the ASR metrics, Chirp~2 and MMS give the most useful round-trip signal, but with nine systems there is not enough statistical power to separate them. The modest $\rho=0.40$ for SCOREQ is worth flagging: signal-level quality alone is an unreliable proxy for what Nepali listeners actually judge.

\begin{table}[t]
\centering
\caption{Agreement between automated metrics and human MOS over the nine systems with automated coverage. CER metrics are ranked with lower CER as better.}
\label{tab:auto_slt}
\footnotesize
\begin{tabular}{lcc}
\toprule
Metric & $\rho$ vs. human MOS & Raw $p$ \\
\midrule
\textbf{NepaliMOS (ours)} & \textbf{0.90} & \textbf{0.001} \\
Chirp~2 CER & 0.60 & 0.09 \\
MMS CER & 0.58 & 0.10 \\
XLS-R Nepali CER & 0.53 & 0.14 \\
SCOREQ & 0.40 & 0.29 \\
NepConformer CER & 0.30 & 0.43 \\
Whisper large-v2 CER & 0.22 & 0.58 \\
Whisper-small CER & 0.17 & 0.67 \\
\bottomrule
\end{tabular}
\end{table}

The divergence is sharpest at the extremes. Proprietary APIs like ElevenLabs should have won: it scores highest on SCOREQ but native listeners did not prefer it. ElevenLabs v3 leads on SCOREQ (rank 1) but places fourth when sorted by human MOS. Piper runs the other direction: SCOREQ ranks it eighth, yet native listeners rated it second. Gemini Flash adds a third pattern. It has the best Whisper-small CER in the panel, and competitive CER under Chirp~2 and MMS too, yet its human MOS still falls below the top systems. A system can be easy for ASR to transcribe while not sounding right to Nepali listeners — and that gap is itself a diagnostic the benchmark is designed to surface.

Whisper large-v3-turbo is a negative finding worth documenting. Among the nine TTS systems, 2\% to 10\% of files hit $\text{CER}>1.0$, with individual cases of Edge TTS Hemkala reaching CER~186 and gTTS reaching CER~1,115. These are illusions rather than transcription errors: the ASR system produces believable Nepali text that is independent of the input audio. Median CER is far lower, so naive per-system averages get pulled upward by rare but extreme hallucinations. We exclude it from the main automated table. For Nepali round-trip work, hallucination filtering is not optional with this model.

NepaliMOS shows that language-matched supervision helps, but the effect has limits that matter in practice. On a random 90/10 utterance split within the dense panel, it reaches $\rho=0.59$ at utterance level and $\rho=0.90$ at system level against human MOS. Those numbers look strong, but they are in-distribution: the model has seen clips from every system in the test set. Leave-one-system-out testing tells a different story. When all clips from one TTS system are removed from training and reserved for testing, utterance-level transfer drops to $\rho=0.43$ and held-out system means are not ranked reliably. LOSO results show that NepaliMOS transfers poorly to unseen providers, and that was not expected. That is the relevant test for anyone wanting to apply NepaliMOS to a new provider. For that use case, a small human calibration set should come first. Uncertainty-aware prediction, attentive pooling, and ranking-aware training objectives are all worth exploring in future work; none of these are part of the current checkpoint.

Post-analysis pilots on the wider 40-system release, including VoxCPM-family models and Dragneel Nepali Whisper ASR checks, were also run. These are not part of the main statistical claims because their rating coverage is sparse and not directly comparable to the dense panel. They are useful for triage. One pattern that shows up repeatedly: unsupported multilingual speech models can produce output that looks fluent in text but is unusable as Nepali audio. The benchmark is designed to surface exactly that kind of divergence.

\section{Conclusion}

NepTTS-Bench is a shared evaluation setting for Nepali TTS: 205 prompts with 77 minimal pairs, audio from 40 named systems or voices, dense MOS data from 180 native speakers across 10 TTS configurations and two human references, ASR round-trip CER from seven models, SCOREQ output, and NepaliMOS. The most valid conclusions that can be drawn from the human data are comparative. Inter-rater agreement at 0.17 shows that native listeners have genuinely varied opinions on TTS. The scores for several current systems fall within confidence intervals of each other, and with Krippendorff's $\alpha=0.17$, this protocol does not have enough resolution to completely separate them.

Automated metrics are useful, but only partially. SCOREQ and ASR-CER can help diagnose certain failure modes and should not be ignored. They are not a replacement for native speaker MOS, and the ASR model you choose affects which systems are ranked where. Of the seven ASR systems we tested, Chirp~2 and MMS matched human scores best. NepConformer is the only Nepali-trained ASR in the set and works well as a baseline, but does not outperform Chirp~2 or MMS here. Whisper large-v3-turbo requires hallucination filtering to be usable on Nepali at all. NepaliMOS has the strongest automatic signal on the dense panel, but LOSO testing shows that human calibration is needed before use with a new provider. The core output is the benchmark package itself. Upon acceptance, it will include Nepali prompts, native speaker ratings, generated audio, an evaluation CLI, and NepaliMOS as a language-matched quality predictor whose limits are measured and reported.

\section*{AI-Generated Content Disclosure}
Generative AI tools were used solely for language polishing and editing during the preparation of this manuscript. These tools assisted with improving clarity, grammar, and conciseness of the writing. No generative AI system was used to generate experimental results, analyses, figures, or scientific conclusions. All technical content, experiments, and interpretations were developed and verified by the authors.

\bibliographystyle{IEEEtranN}
\bibliography{references}

\end{document}